# A Prototype of Trigger Electronics for LAWCA Experiment[*]

YAO Lin(姚麟) [1,2] ZHAO Lei(赵雷) [1,2;1)]
SHANG Lin-Feng(商林峰) [1,2]   LIU Shu-Bin(刘树彬) [1,2]    AN Qi(安琪) [1,2]

[1] State Key Laboratory of Particle Detection and Electronics, University of Science and Technology of China, Hefei, 230026, China
[2] Department of Modern Physics, University of Science and Technology of China, Hefei, 230026, China

**Abstract:** The Large Area Water Cherenkov Array (LAWCA) experiment focuses on high energy gamma astronomy between 100 GeV and 30 TeV. Invoked by the idea of hardware triggerless structure, a prototype of LAWCA trigger electronics is implemented in one single VME-9U module which obtains all the data from the 100 Front End Electronic (FEE) endpoints. Since the trigger electronics accumulates all the information, the flexibility of trigger processing can be improved. Meanwhile, the dedicated hardware trigger signals which are fed back to front end are eliminated; this leads to a system with better simplicity and stability. To accommodate the 5.4 Gbps system average data rate, the fiber based high speed serial data transmission is adopted. Based on the logic design in one single FPGA device, real-time trigger processing is achieved; the reprogrammable feature of the FPGA device renders a reconfigurable structure of trigger electronics. Simulation and initial testing results indicate that the trigger electronics prototype functions well.

**Key words**: LAWCA, gamma astronomy, PMT, FPGA, trigger electronics

**PACS**: 84.30.-r, 07.05.Hd

## 1 Introduction

The Large Area Water Cherenkov Array (LAWCA) aims at the physical target of gamma astronomy at high energies. Its detector structure and function is identical to the Water Cherenkov Detector Array (WCDA) in the Large High Altitude Air Shower Observatory (LHAASO) project [1 2], with a detector scale equivalent to one fourth part of the latter. The LAWCA detector consists of 900 Photon Multiplier Tubes (PMTs) which are scattered in a $150 \times 150$ m$^2$ area. The detector array is partitioned by curtains into $30 \times 30$ detector units, each with a side length of 5 meters. Each detector unit contains one PMT at the bottom center to collect Cherenkov light generated by the disseminating charged particles in water. The PMTs are installed 4 meters under water and are isolated from each other by black plastic curtains to prevent crosstalk from lights yielded in neighboring cells. Every $3 \times 3$ PMTs form a group and their outputs are fed to one Front End Electronic (FEE) module for time and charge information measurement [3]. Fig. 1 shows the geometric structure of the LAWCA detectors.

[*] Supported by Knowledge Innovation Program of the Chinese Academy of Sciences (KJCX2-YW-N27) and National Natural Science Foundation of China (11175174, 11005107)
1) Email: zlei@ustc.edu.cn

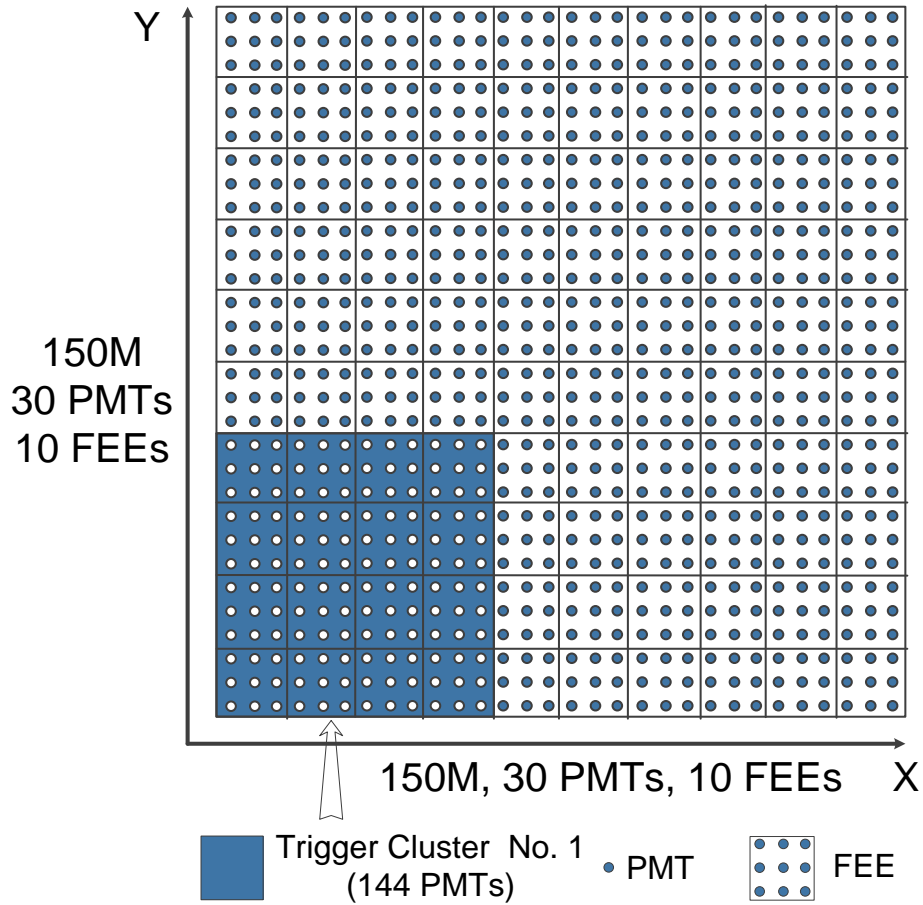

Fig. 1. Geometric structure of the LAWCA detectors.

In traditional trigger electronics such as in AMANDA [4] and in BES III [5], each FEE module sends its hit signals to the main trigger module. After trigger processing, global trigger signals will be sent back to FEEs for data readout. To improve the flexibility of trigger electronics, some experiments have adopted the "Triggerless" architecture. Typically, this architecture obtains all the data from the FEEs and the kernel trigger processing is done in DAQ (usually a computer farm) [6]. It could achieve a high flexibility but also require much more resources cost on data storage and processing.

Due to the complicated phenomena of cosmic ray showers, a high flexibility of trigger electronics is required to accommodate different potential trigger patterns for better trigger efficiency. Invoked by the idea of hardware triggerless architecture and based on the improvement of data transmission and FPGA techniques, a prototype of LAWCA trigger electronics is designed with all data transferred to a trigger module for processing. Compared with traditional trigger electronics receiving only part of the data from front end, much more detailed information can be utilized to achieve a better trigger flexibility. It also eliminates the need of trigger signals fed back to front end; thus a better system simplicity can be accomplished. In the trigger module, one single FPGA device is employed to carry out the real-time trigger processing. Based on the abundant programmable inner connections and logic resources in the FPGA, reconfigurable trigger electronics can be realized; meanwhile, the trigger algorithms can be upgraded online with future modifications. Besides, because a large part of fake events can be rejected by the trigger electronics, the requirement on DAQ is significantly reduced.

The structure of the overall LAWCA readout electronics is shown in Fig. 2. The readout electronics is synchronized under a global high precision 40 MHz clock. This clock is distributed based on a simplified White Rabbit protocol [7]. 100 FEEs are adopted to digitize the signals from the 900 PMTs, and there are 10 Clock & Data Transmission Modules (CDTMs) to accumulate the data and distribute the global clock to the FEEs. All the data from the CDTMs will be transferred through fibers to the trigger module for processing.

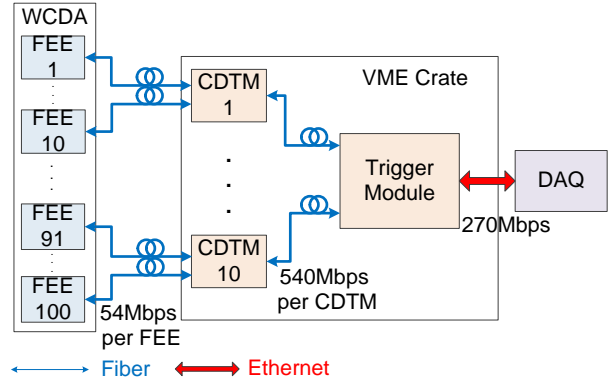

Fig. 2. Structure of the LAWCA readout electronics.

In this paper, the LAWCA trigger electronics prototype and initial testing results are presented. Section 2 introduces the requirement of the trigger electronics. Section 3 describes the structure of the trigger prototype in detail. Section 4 presents the initial testing results. Finally, Section 5 gives a summarized conclusion.

## 2 Requirement of the LAWCA Trigger Electronics

### 2.1 Basic Trigger Pattern

As for the 900 PMTs in LAWCA, each has its individual signal readout channel which is responsible for both time and charge measurement. Every $3 \times 3$ channels are integrated in one FEE. The photo-electron number (nPE) measuring range starts from 1 PE to 4000 PE. The raw event rate of each PMT is estimated to be around 50 kHz, which contains valid signals and noises. The pulse width is an important reference for the design of trigger pattern. As for valid signals, simulations show that the pulse width is mostly smaller than 13 ns. The pulse width turns larger for slant showers, but is still smaller than 18 ns [8].

As in previous statement, there may exist different trigger patterns in the future experiment. This trigger electronics prototype is designed and evaluated based on a basic trigger pattern as follows. Considering the shower area of one cosmic ray secondary, the PMTs are divided into 16 overlapped trigger clusters, and each cluster is a $12 \times 12$ PMT array. The X axis view of the overlapped trigger clusters' locations is shown in Fig. 3; the view in the Y axis is quite similar.

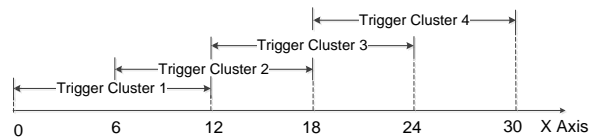

Fig. 3. X axis view of the trigger clusters.

When no less than 12 PMTs in any trigger cluster are fired within any 250 ns duration, it will be recognized as a valid event and a global trigger signal will be generated. To obtain more detailed information of one shower, a 2 μs readout time window is demanded. Therefore, the complete data section of one shower includes the data recorded both before and after 1 μs of the trigger signal.

**2.2 Data structure and transmission requirement**

The charge and time information from the fired PMTs is recorded by FEEs. With each fired PMT, the corresponding FEE produces a 96-bit data packet. As shown in Fig. 4, it contains a 3-bit packet header, 3-bit checksum, 10-bit channel number, 32-bit time information, 32-bit charge information and 16 bits reserved for future extension. The header and checksum are used for packet identification and transmission error detection. As for the 32-bit time information, the first part is for the course time, which is the output of a 27-bit counter at the frequency of 40 MHz. The second part is the fine time result of a TDC with a 1 ns bin size. It is 5-bit wide to cover the 25 ns global clock cycle. To cover the charge dynamic range from 1 PE to 4000 PE, two measurement channels with different gains are employed in the FEEs. The high gain channel responds for the range from 1 PE to 133 PE, while the low gain channel responds for the range from 133 PE to 4000 PE. Each channel records 16-bit charge information and thus the total data width is 32 bits. In this 96-bit data packet, the channel number and time information is extracted for trigger processing.

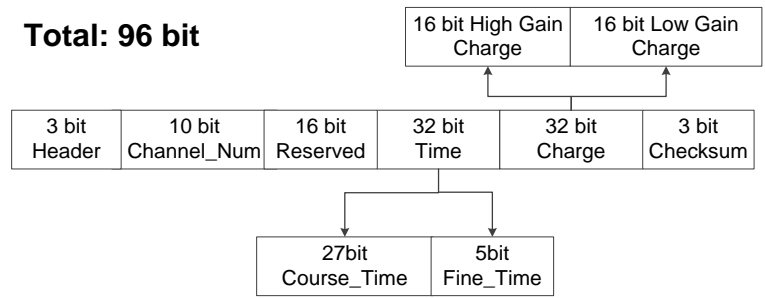

Fig. 4. FEE data packet structure.

Since the raw event rate is about 50 kHz and each FEE contains 9 PMT channels, the average raw data rate of one FEE is about 9 × 96 bit × 50 kHz = 43.2 Mbps. To achieve clock distribution along with data transmission in a large area, the fiber based serial transmission method is adopted. To reduce transmission loss in long distance serial transmission, the 8B/10B encoding is adopted [9]. It will import an extra 25% transmission cost but greatly reduce the bit error probability. Therefore, the final average data rate of one FEE is about 54 Mbps. Data from 10 FEEs will be collected by one CDTM and all the data from frond end will be transferred to the trigger module in fibers through the 10 CDTMs at a 5.4 Gbps data rate.

As mentioned above, the trigger electronics of LAWCA is responsible for the input 5.4 Gbps data buffering, trigger processing and corresponding valid data selection with a 2 μs readout window, as well as communication with DAQ.

**3 Design of the trigger electronics prototype**

**3.1 Hardware architecture**

As previously presented, invoked by the hardware triggerless idea [6 10], in the LAWCA readout electronics all the data from FEEs are transferred directly to the trigger module for selection without the need of trigger signals feedback wires. Therefore, a better flexibility and system simplicity can be guaranteed. With the input 5.4 Gbps data stream, the main challenge is how to realize the real-time trigger processing with such a high data rate. When facing similar difficulty, the PANDA experiment adopted a full-mesh compute node network based on an FPGA array. In a single compute node, five Xilinx Virtex-4 FX60 FPGAs are used to execute the algorithms with a 2 Gbps data rate. The FPGAs communicate with each other in 32 parallel lines routed on the Printed Circuit Board (PCB). The total data rate in LAWCA is much less than in PANDA (200 GBps) [11]. With recent development of data transmission and FPGA technique, the LAWCA trigger electronics prototype can be implemented with a higher integration. One single FPGA (Xilinx Virtex-6 xc6vlx240t) is adopted to implement the kernel trigger processing algorithms. Compared with the FPGA array scheme, it requires less power consumption and provides reconfigurable inner connections which guarantee a much higher flexibility. This FPGA integrates 24 internal serializer/deserializer GTX interfaces, each with a maximum transmission rate up to 6.75 Gbps [12]. 10 fibers with 1.25 Gbps SFP connectors are adopted, supporting a total data rate up to 12.5 Gbps. Of course, the data transmission speed could easily be further enhanced with faster SFP connectors.

The LAWCA trigger electronics prototype is implemented in a standard VME-9U module. To accommodate the potential upgrade of trigger processing algorithms, the online reconfiguration method [13] is adopted. The architecture of the trigger module is shown in Fig. 5. It consists of 4 major parts -- Data Preprocessing, Data Buffer, Trigger Processing and Data Transmission. The Data Preprocessing part receives the 5.4 Gbps data stream from the 10 fibers, and then it extracts the hit flags which are 1-bit binary numbers to identify whether the corresponding PMT is fired. The raw data stream will be stored in the Data Buffer and wait for trigger processing result. The Trigger Processing part executes the trigger algorithm using the hit flags and generates a global trigger signal when the trigger condition is met. With this global trigger, a 2 μs valid data section will be selected from the Data Buffer and then transferred through the Data Transmission part to DAQ.

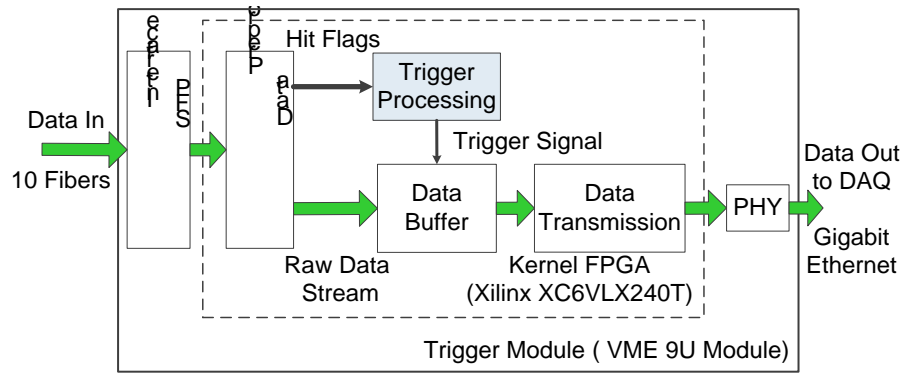

Fig. 5. Block diagram of the LAWCA trigger module.

**3.2 Data stream Synchronization**

As previously mentioned, the average data rate is 5.4 Gbps. In fact, there exists the possibility that one shower of cosmic ray secondary would cause part of the FEEs to generate a much higher data rate than the others in a short time period. In this situation, it is possible that the data transmission in fiber A is quite busy while very little data is under transmission in fiber B. This unbalance may cause the un-alignment of the events' data packets generated in the same clock cycle. Thus great care should be taken for the data alignment/synchronization among different channels.

The synchronization of the data streams is implemented with a delay method. First, all the data received by the trigger module are buffered, and then each data packet is loaded for trigger processing after a determined time interval since it is generated (the time point is recorded in its packet). Since the system data transmission capacity is higher than the average data generation rate, the unbalance of data transmission in different fibers can be reduced with this determined interval.

In fact, the data transmission unbalance is caused by the different amounts of fired PMTs in different CDTMs. The most severe unbalance would occur when many PMTs in one CDTM are fired while other CDTMs have no PMT fired.

Simulations indicate that the possibility of more than 40 PMTs in one CDTM fired at the same time is less than $2.68 \times 10^{-15}$. Synchronization with up to 40 PMTs fired coincidentally in one CDTM is acceptable for the LAWCA trigger electronics. With the 1.25 Gbps data transmission rate of one fiber channel, it will cost 3.84 μs to transfer all data of the 40 fired PMTs. Therefore, a 4 μs delay buffer is adopted to avoid data loss, which is equivalent to 160 cycles of the 40 MHz global clock.

### 3.3 Design of the hit flag array

According to the trigger pattern, whether a PMT is fired is one of the judgment conditions for the global trigger decision. Hit flags are adopted to represent the PMT status in the trigger logic. The hit flag is a 1-bit binary number, and thus a 900 × 172 array is required to storage the 900 channels' hit flags, as shown in Fig. 6. The column number 172 corresponds to a time delay of 4.3 μs with the clock frequency of 40 MHz (0.3 μs is reserved as the hit flag margin). In every data packet received by the trigger module, the hit flag will be extracted and written in its corresponding position of the array according to its 10-bit address number and its course time information (mod by 172).

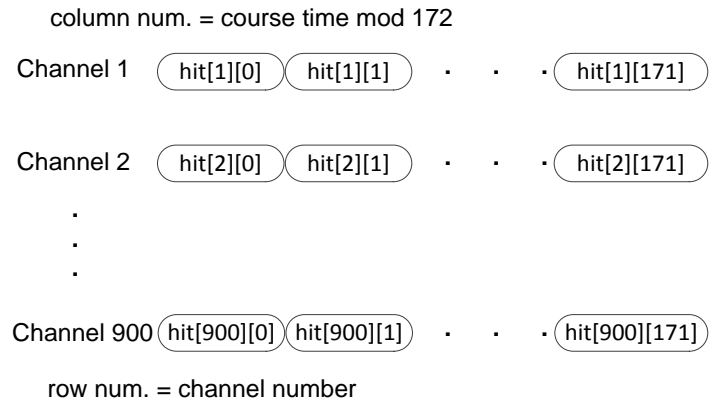

Fig. 6. Structure of the hit flag array.

When a PMT is fired, the count of other PMTs fired in the following 250 ns window is necessary for trigger. Since the hit flag exists for only 1 clock cycle, the digital pulse broadening is required. When a data packet is received, the corresponding hit flag and the following 9 hit flags in the same row will be assigned to 1. These 10 flags cover 10 clock cycles which is equivalent to the time duration of 250 ns.

The trigger processing is conducted in every clock cycle, which requires real-time refreshing of the hit flag array. Therefore, this array is designed in a ring structure with end to end closure. Fig. 7 depicts the 172 hit flags arrangement in PMT channel No. 1 while the other 899 channels are organized in the same structure. At clock cycle N, the hit flag under processing is hit[1][N] (marked by the Trigger Pointer) while the possible location of the Write Pointer is in an active section that covers the 160 positions from N + 1 to N − 12 in Fig. 7 (this corresponds to a 4 μs delay buffering). The 10 positions from N − 2 to N − 11 are reserved as the hit flag margin. A clear pointer is adopted to achieve refreshing of the hit flag array, and it points at position N − 1 where the flag is already used in trigger processing at the previous cycle.

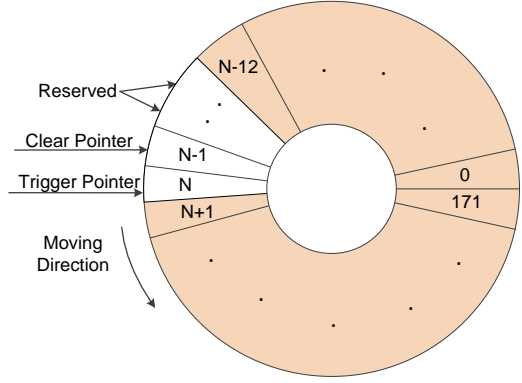

Fig. 7. The hit flag cycling structure in PMT channel No.1.

**3.4 Kernel trigger processing**

In the trigger pattern mentioned in Section II, if no less than 12 PMTs are fired in any cluster within any 250ns duration, a trigger decision will be made with a global trigger signal pulse generated. In fact, this global trigger (denoted as Trigger_G) is the combination of local trigger signals of the individual trigger clusters, which is denoted as Trig_m (m: 1~16). The relationship can be expressed as:

$$\text{Trigger\_G} = (\text{Trig\_1}) \parallel (\text{Trig\_2}) \parallel \ldots\ldots \parallel (\text{Trig\_16}) \quad (1)$$

In every clock cycle, the 900 hit flags in one column marked by the Trigger_Pointer are loaded for trigger processing. Corresponding to the trigger cluster structure, these 900 hit flags are divided into 16 overlapped groups and local trigger processing is conducted separately. Since the local trigger processing of all the 16 trigger clusters is identical, we take the local trigger algorithm in cluster No. 1 as an example, which ranges from Channel 1 to Channel 144. As shown in Fig. 8, the number of fired PMTs in the current clock cycle (denoted as Trig_1_cnt) can be expressed as:

$$\text{Trig\_1\_cnt} = \sum_{k=1}^{144} \text{Hit\_1\_k} \quad (2)$$

where Hit_1_k refers to the hit flag in PMT channel No. k.

Fig.8. Hit flags of trigger cluster No.1 for local trigger processing.

If Trig_1_cnt is no less than 12, a local trigger signal Trig_1 will be generated, and then Trigger_G is asserted to start valid data selection and readout. In the next clock cycle, the 144 hit flags in the next column will be loaded for processing and the current column will be cleared. Based on this scheme, the kernel trigger processing is conducted continuously.

In the FPGA implementation of the trigger algorithm, a simplified 16 × 172 array is used to replace the 900 × 172 hit flag array. Each element in the simplified array directly records the calculation result Trig_m_cnt (m: 1~16), which greatly reduces the logic resource cost.

### 3.5 Data buffering

As previously described, the incoming data will be stored in the Data Buffer while the hit flags are extracted for trigger processing. Considering the requirement of 4 μs delay buffering and a 2 μs readout window, the Data Buffer is divided into 7 slices, each containing the data within a 1 μs time period. There are Buffering Pointer, Processing Pointer, Readout Pointer and Cleanup Pointer to indicate the correct addresses for corresponding data operations. The determined time relationships between the slices and pointers are shown in Fig. 9.

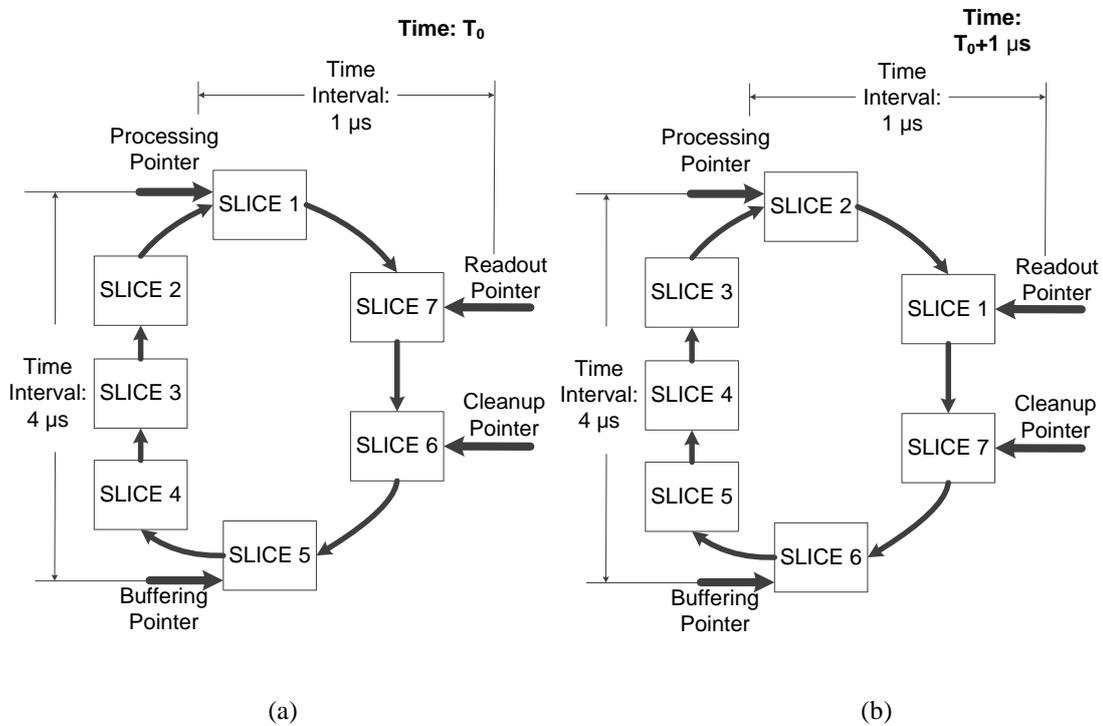

Fig. 9. Organization of the 7 buffer slices at (a) $T_0$ (b) $T_0+1$ μs.

The Buffering Pointer indicates which slice is buffering the current data from the Data Preprocessing part. The Processing Pointer indicates the slice which contains the data corresponding to the hit flags currently under trigger processing. Considering the 4 μs delay, its location is 4 slices behind the Buffering Pointer. The Readout Pointer refers to the starting slice for valid data readout, which is 1 μs behind the Processing Pointer. When a global trigger signal is generated, the data of 3 slices (Slice 7, 1 and 2 in Fig. 9(a)) will be read out. The Cleanup Pointer indicates that the data in this slice is obsolete and can be cleared. Similar to the hit flag array, the slices are organized in a ring structure and the pointers will shift to the next slice every 1 μs, as shown in Fig. 9.

### 3.6 Data transfer interface

After data selection, the valid data will be packaged and transferred to DAQ through the Gigabit Ethernet. The Gigabit Ethernet interface is implemented with an FPGA internal MAC Intellectual Property (IP) core and an external PHY chip. Detailed implementation can be referred in [14].

## 4 Initial testing results

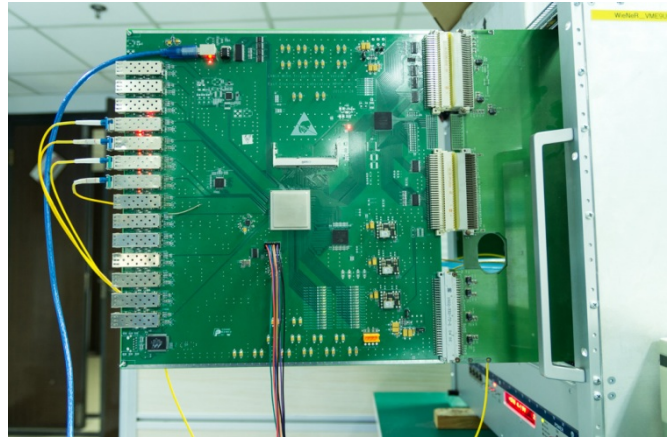

Fig. 10. Trigger module under test.

The LAWCA trigger electronics prototype has been designed and initial tests have been conducted. The key capability of the trigger electronics is fiber based high-speed data transmission and real time trigger processing. Eye diagram measurement of the 1.25 Gbps data transmission between the CDTM and the trigger module was conducted. It was measured in the Waveform Database (WfmDB) mode [15] of the Tektronix DPO 7354C Digital Phosphor Oscilloscope. The recorded waveform is shown in Fig. 11; the measured timing margin is around 700 ps and the amplitude margin is about 290 mV.

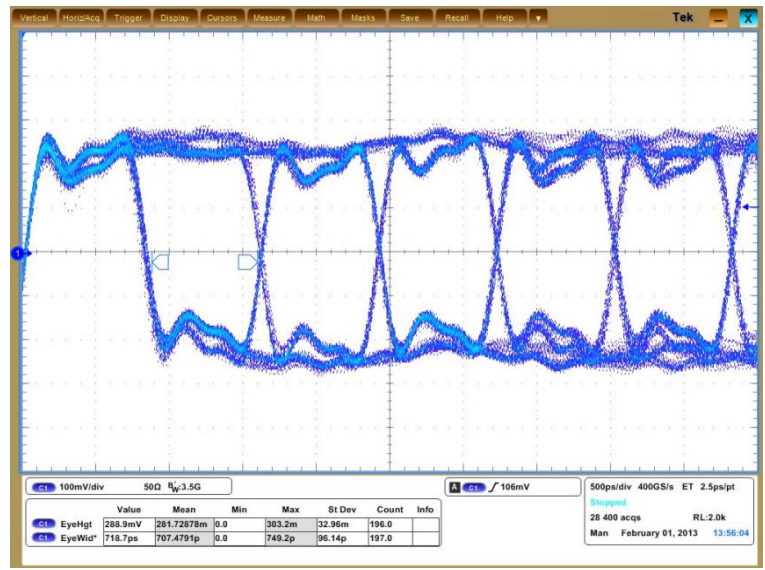

Fig. 11. Eye diagram test result.

Since FEEs and detectors R&D is still in progress, simulation data of PMT behavior under air shower is imported to evaluate the correctness of the trigger logic in the FPGA. The simulation data is generated by physicists with the Geant4 and CORSIKA software. It contains the charge and time information from PMT signals caused by cosmic ray secondary or background noises. 106 simulations are conducted in parallel and information of the total 23803 events is recorded. The simulation results are used as the data source, which is transferred to the trigger module to be processed by the trigger logic in the FPGA, and then the valid data are obtained through the Ethernet interface. We also carried out MATLAB modeling based on the trigger pattern, with which the same simulation data is processed. The results of the two processes are compared. Fig.12 shows the valid data ratio at the output port of our prototype trigger module and of the MATLAB platform. The results are exactly matched, which proves that the trigger prototype functions well with the trigger logic strictly following the expected trigger pattern.

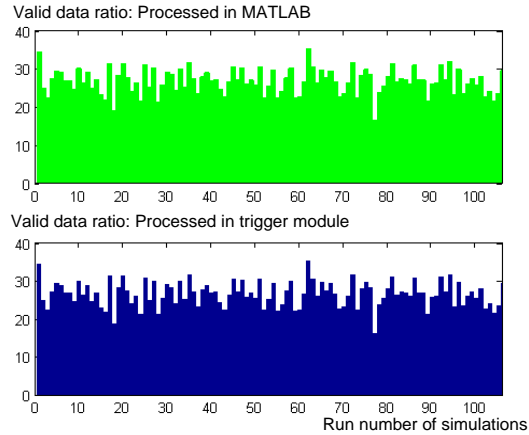

Fig. 12. Valid data ratios calculated in MATLAB and the trigger module.

With the current trigger pattern, the logic resource consumption in the FPGA is estimated. As shown in Table 1, the occupied slices are up to 65%. The trigger logic could be revised with upgraded trigger patterns. Other FPGA devices with larger capacity (such as xc6vlx365t and xc6vlx550t) can be employed to accommodate potential more complicated trigger algorithms.

Table. 1. Logic slices consumption of the FPGA.

| Slice Logic Utilization | Used | Available | Utilization |
|---|---|---|---|
| Number of Slice Registers | 11,566 | 301,440 | 3% |
| Number of Slice LUTs | 87,241 | 150,720 | 57% |
| Number of occupied Slices | 24,545 | 37,680 | 65% |

## 5 Conclusion

In this paper, a prototype of the LAWCA trigger electronics and corresponding initial testing results are presented. This trigger prototype can achieve trigger processing and data selection of the 900 PMTs at an average data rate of 5.4 Gbps. To deal with such a high data rate, the fiber based serial data transmission method is adopted. With all information accumulated, the trigger electronics could achieve a high trigger flexibility; with no trigger signal fed back to front end, it could accomplish a high simplicity. All trigger algorithms are implemented in one single FPGA device. With the reprogrammable feature of the FPGA, a reconfigurable structure of trigger electronics can be guaranteed. Initial testing results indicate that this trigger electronics prototype functions well, which also provides a good technical foundation for the future LHAASO experiment.

**Acknowledgement**

This work is supported by the Knowledge Innovation Program of the Chinese Academy of Sciences (KJCX2-YW-N27) and the National Natural Science Foundation of China (No. 11175174 and No.11005107). The authors would like to appreciate all of the LAWCA collaborators who helped to make this work possible.

———————————————